\documentclass[prc,twocolumn,showpacs,preprintnumbers,amsmath,amssymb]{revtex4}
\usepackage{graphicx}
\begin{document}

\title{ Self-consistent many-body approach to the electroproduction of hypernuclei}
\author{P. Byd\v{z}ovsk\'y$^a\footnote{bydzovsky@ujf.cas.cz}$, D. Denisova$^{a,b}$, D. Petrellis$^a$, D. Skoupil$^a$, P. Vesel\'y$^a$, \\ G. De Gregorio$^{c,d}$, F. Knapp$^{b}$, and N. Lo Iudice$^{e}$}
\affiliation{$^a$Nuclear Physics Institute, ASCR, 25068 \v{R}e\v{z}/Prague, Czech Republic \\
$^b$Institute of Particle and Nuclear Physics, Faculty of Mathematics and Physics, \\ Charles University, V Hole\v{s}ovi\v{c}k\'{a}ch 2, 180 00 Prague, Czech Republic \\
$^c$Dipartimento di Matematica e Fisica, Universit$\grave{a}$ degli Studi della \\ 
Campania "Luigi Vanvitelli", viale Abramo Lincoln 5, I-81 100 Caserta, Italy \\
$^d$ Istituto Nazionale di Fisica Nucleare, Complesso Universitario di Monte S. Angelo, Via Cintia, I-80 126 Napoli, Italy \\
$^e$ Dipartimento di Fisica, Universit$\grave{a}$ di Napoli Federico II, 80126 Napoli, Italy}

\date{\today}

\begin{abstract}
The electroproduction of selected $p$- and $sd$-shell hypernuclei was studied within a many-body approach using realistic interactions between the constituent baryons. 
The cross sections were computed in the distorted-wave impulse approximation using two elementary amplitudes for the electroproduction of the $\Lambda$ hyperon. 
The structure of the hypernuclei was investigated within the framework of the self-consistent $\Lambda$-nucleon Tamm-Dancoff approach and its extension known as the $\Lambda$-nucleon equation of motion phonon method. 
Use was made of the NNLO$_{\rm{sat}}$ chiral potential plus the effective Nijmegen-F $YN$ interaction.  The method was first implemented on light nuclei for studying the available experimental data and establishing a relation to other approaches. After this proof test, it was adopted for predicting the electroproduction cross section of the hypernuclei $^{40}_{~\Lambda}$K and $^{48}_{~\Lambda}$K in view of the E12-15-008 experiment in preparation at JLab. 
On the ground of these predictions, appreciable  effects on the spectra are expected to be induced by the $YN$ interaction.   
 \end{abstract}

\maketitle
%
\section{Introduction}
Hypernuclei  represent an important bridge between particle and nuclear physics. Their study is of great relevance to both hadron physics and nuclear structure \cite{GHM2016,topics}.

Their spectra offer a unique tool for studying the effective hyperon-nucleon ($YN$) interaction, especially its spin-dependent component which is  not clearly deduced from scattering experiments~\cite{HT2006,HMY2010,JohnM}. 
The replacement of a nucleon with the $\Lambda$ hyperon also allows one to investigate deeply-bound states of $\Lambda$ as the hyperon is not Pauli blocked inside the hypernucleus. 
It is especially important to investigate these deeply-bound states in heavy hypernuclei with a large neutron excess, e.g. in $^{208}_{~~\Lambda}$Tl~\cite{208Pb}, as the interior of such heavy systems is a good proxy of nuclear matter. 
The comparison of the hypernuclear spectra with the experimental data contributes to clarification of  the reaction mechanism and nuclear structure models and offers an additional test-ground for the nuclear forces.

Among the reactions used to produce hypernuclei~\cite{GHM2016,HT2006}, the electroproduction is of particular interest since the electromagnetic interaction is well known and can be treated perturbatively. 
In the assumed kinematics, with large ($\approx 1$ GeV) photon and kaon momenta, the reaction can be satisfactorily described in the distorted-wave impulse approximation (DWIA)~\cite{fermi,HYP2018,AP2019}.   
Moreover, since the energy resolution achieved in experiments with electron beams is better than that  in hadron induced reactions~\cite{HT2006}, it is possible to extract more precise information about the reaction and the details of the effective $YN$ interaction. 

In order to analyze the data unambiguously one needs a reliable theoretical model where  inherent approximations are clearly specified as well as the input information (elementary amplitude, structure calculations, and kaon distortion).  
Such a need emerged from a recent investigation \cite{fermi} where the importance of the Fermi motion and other kinematic effects on the electroproduction of hypernuclei, induced by two elementary amplitudes, was pointed out thereby suggesting some limits of the predictive power of the existing DWIA calculations. 

The Fermi motion effects were also investigated by Mart {\it et al.}~\cite{Mart1,Mart2} in the electromagnetic production of the hypertriton utilizing the two-component form of the elementary production amplitude in a general reference frame~\cite{PB}. 
Using the KAON-MAID~\cite{KM} model for the elementary amplitude, Mart and  Ventel~\cite{Mart2} found that the Fermi motion effects are essential for a correct description of the hypertriton electroproduction. These effects were found to be sizable in the longitudinal part of the cross section in agreement with our recent conclusions made for heavier hypernuclei~\cite{fermi}. 
Therefore, in this  work we will use the optimum on-shell approximation discussed in Ref.~\cite{fermi} which partially includes the Fermi motion via the proton optimum momentum thereby allowing the use of the on-energy-shell elementary amplitude.

In previous papers \cite{HYP2018,AP2019,fermi}, we have investigated the cross section for the electroproduction of $p$-shell hypernuclei  within a phenomenological shell-model, whose nucleon-nucleon ($NN$) and effective $YN$ interactions were fitted to very precise data from $\gamma$-ray spectroscopy~\cite{JohnM}. 

In the present work we will use the $p$- and $sd$-shell hypernuclear space and investigate, in addition to the light systems, the medium-mass hypernuclei   $^{28}_{~\Lambda}\text{Al}$,  $^{40}_{~\Lambda}$K, and $^{48}_{~\Lambda}$K. 
The electroproduction of the latter two hypernuclei will be measured in the E12-15-008 experiment planned at JLab~\cite{E12-15-008,Nue}. 

The  excitation spectra  of $p$-shell and $sd$-shell $\Lambda$-hypernuclei were investigated within several shell-model approaches using different hypernuclear wave functions \cite{Motoba1994,Sotona1994,Motoba2010,Motoba2012,Motoba2017}. 
In Ref. \cite{Motoba2012}, the full $(sd)^n$-space  and the Saclay-Lyon A (SLA) elementary amplitude in the frozen-proton approximation were adopted to investigate the photoproduction of the medium-mass hypernuclei $^{28}_{~\Lambda}$Al and $^{40}_{~\Lambda}$K.
In Ref.~\cite{Cohen}, a simple elementary-production amplitude and particle-hole shell model configurations were used for determining the energy-dependent cross sections of the electroproduction of $^{40}_{~\Lambda}$K.  
Recently, the antisymmetrized molecular dynamics was adopted for investigating the structure of light $\Lambda$-hypernuclei and their spin dependence through  the $\Lambda N$ interaction~\cite{Isaca2020,Motoba2022}.
The shell-model was also utilized for predicting the cross section of the photoproduction of $^{12}_{~\Lambda}$B  \cite{Lee1998}. 
A fully covariant model was employed in description of the photoproduction of $^{16}_{~\Lambda}$N \cite{Lenske2008}. In this calculation the  nucleon and hyperon bound states were obtained by solving the Dirac equation with a static nuclear mean-field potential. The same approach was also adopted for describing the hadron-induced reactions~\cite{Shyam2008}. 
In Refs.~\cite{Benn1989,Benn1987} the photoproduction of $^{16}_{~\Lambda}$N and $^{12}_{~\Lambda}$B was studied in the relativistic distorted wave impulse approximation. Here the single particle (s.p.) bound states were solutions of the time-independent Dirac equation using scalar and vector potentials~\cite{Benn1989}.

The study presented here is carried out within a microscopic self-consistent many-body approach which involves complex nuclear excitations.
A Hartree-Fock (HF) s.p. basis is generated for $\Lambda$ and nucleons from the effective $YN$ interaction plus the chiral NNLO$_{\rm{sat}}$ $NN$ + $NNN$ potential \cite{NNLOsat}. 
The  residual Hamiltonian is diagonalized in the particle-hole  basis $\{\mid p-h \rangle\}$   so obtained. This mean field approach, known as Tamm-Dancoff for hypernuclei (TD$_\Lambda$), was adopted for $p$-shell hypernuclei in Ref.~\cite{Ves3}.

Here, we show how to go beyond the mean field theories and extend the TD$_\Lambda$  by coupling the $\Lambda$-nucleon $p-h$ states to more complex excitations of the nuclear core. Such a goal is reached within the equation of motion phonon method (EMPM) applied to hypernuclei.
 
The nuclear EMPM, in its  upgraded version \cite{Bianco12}, adopts the equations of motion to generate an orthonormal basis of $n$-phonon ($n=0,1,2...$) states. Such a basis is then adopted to diagonalize the residual Hamiltonian. This amounts to coupling the  TD  $p-h$ configurations  to $np-nh$ states. The  method  can use any Hamiltonian and does not rely on approximations except for the truncation of the multiphonon space.
It has been used for investigating bulk and spectroscopic properties of light as well as heavy nuclei~\cite{EMPM1a,EMPM2,EMPM4,EMPM8}.
 
The formulation of the method for hypernuclei,  which we call EMPM$_{\Lambda}$, has two variants. One couples the $\Lambda$ particle to an even-even nuclear core~\cite{Ves5}. This version is suitable for describing hypernuclei such as $^{5}_{\Lambda}$He, $^{17}_{~\Lambda}$O, and $^{41}_{~\Lambda}$Ca. The other variant describes the hypernuclear states as $\Lambda$-$N$ particle-hole excitations of an even core~\cite{Ves6,Ves7} within the TD${_\Lambda}$ framework. This version is suitable for describing  hypernuclei like  $^{4}_{\Lambda}$H, $^{16}_{~\Lambda}$N, $^{40}_{~\Lambda}$K, and $^{48}_{~\Lambda}$K.   
It is worth pointing out that the calculation presented here is self-consistent and parameter free. 
 
The paper is organized as follows: Sec. II briefly introduces the formalism developed in  Ref.~\cite{fermi} and used here to compute the cross sections in the DWIA. 
Sec. III deals with the TD$_\Lambda$ and EMPM$_\Lambda$ methods describing the structure of the hypernuclei. Calculation details are discussed in Sec. IV. In Sec. V, we first present the excitation spectra of the $p$-shell hypernuclei $^{12}_{~\Lambda}$B and $^{16}_{~\Lambda}$N and compare them with the experimental data as well as with our advanced shell model results~\cite{AP2019,fermi}. 
We then discuss the $sd$-shell hypernuclei $^{28}_{~\Lambda}$Al, $^{40}_{~\Lambda}$K, and $^{48}_{~\Lambda}$K and give predictions for the planned JLab experiment E12-15-008. 
The conclusions drawn from our study are given in Sec. VI.
%
%
 \section{The cross section}
The model for computing  the cross section of the electroproduction of hypernuclei associated with a kaon in the final state 
\begin{equation}
e + A \longrightarrow e^\prime + H + K^+
\label{electro}
\end{equation}
is described in Ref. \cite{fermi}. 

The triple differential cross section in the nucleus-rest frame is 
\begin{equation}
\frac{d^3\sigma}{dE_e^\prime d\Omega_e^\prime d\Omega_K} = 
\Gamma\,\frac{d\sigma}{d\Omega_K},
\end{equation}
where
\begin{equation}
\frac{d\sigma}{d\Omega_K} = 
\frac{d\sigma_{\sf T}}{d\Omega_K} +
\varepsilon_{\sf L} \frac{d\sigma_{\sf L}}{d\Omega_K} +
\varepsilon \frac{d\sigma_{\sf TT}}{d\Omega_K} +
\sqrt{\varepsilon_{\sf L}(\varepsilon+1)}\;\frac{d\sigma_{\sf TL}}{d\Omega_K}.
\label{crs}
\end{equation}  

Here $\varepsilon_{\sf L}$ and $\varepsilon$ are the longitudinal and transverse virtual-photon polarizations, respectively, and $\Gamma$ is the virtual photon flux~\cite{fermi}. The separate cross sections are  
\begin{equation} 
\frac{d\sigma_{\sf T}}{d\Omega_K} = \frac{\beta}{2[J_A]^2}\sum_{Jm}
\frac{1}{[J]^2}\left(\,|A^{+1}_{Jm}|^2 + |A^{-1}_{Jm}|^2\right)\,,
\label{crs1}
\end{equation}
\begin{equation} 
\frac{d\sigma_{\sf L}}{d\Omega_K} = \frac{\beta}{[J_A]^2}\sum_{Jm}
\frac{1}{[J]^2}|A^0_{Jm}|^2\,,
\label{crs2}
\end{equation}
\begin{equation} 
\frac{d\sigma_{\sf TT}}{d\Omega_K} = \frac{\beta}{[J_A]^2}\sum_{Jm}
\frac{1}{[J]^2}{\sf Re\,}[A^{+1}_{Jm}A^{-1*}_{Jm}] \,,
\label{crs3}
\end{equation}
\begin{equation} 
\frac{d\sigma_{\sf TL}}{d\Omega_K} = \frac{\beta}{[J_A]^2}\sum_{Jm}
\frac{1}{[J]^2}{\sf Re\,}[A^{0*}_{Jm}(A^{+1}_{Jm} -A^{-1}_{Jm})],
\label{crs4}
\end{equation}
where $[J] = \sqrt{2J + 1}$ and $\beta$ is the kinematical factor~\cite{fermi}. The transverse part (\ref{crs1}) corresponds to the photoproduction cross section. 

The reduced amplitudes $A^\mu_{Jm}$ have the form
\begin{equation}
A^\mu_{Jm} =\frac{1}{[J]}  
\sum_{rs}  {\cal W}^\mu_{Jm} (rs)
(\Psi_H\,||\,(b_{r}^\dagger\times a_s)^J\,||\;\Phi_A\,),
\label{amplitude-3}
\end{equation}
where 
\begin{equation}
{\cal W}^\mu_{Jm} (rs)= \sum_{LS} {\cal H}^{LSJ}_{l'j'lj}
( \,{\cal R}^{L}_{ (rs)} \times {\cal F}_\mu^S )^{Jm}. 
\label{amplitude-3a}
\end{equation}
Here, the label $\mu$ denotes the virtual photon helicity, $S$ the spin transfer, and $L$ the orbital momentum in the photon-kaon system. ${\cal H}^{LSJ}_{l'j'lj}$ is a geometrical factor  \cite{fermi} which includes 3j and 9j symbols, ${\cal F}^{S\eta}_{\mu}$ the elementary production amplitude, and ${\cal R}^{LM}_{(rs)}$  the radial integral which includes the kaon distortion and the proton ($s$) and $\Lambda$ ($r$) s.p. wave functions.  
Here the radial integrals are calculated using the HF s.p. wave functions under the assumptions described in Ref.~\cite{fermi} for the kaon distortion, consistently with the hypernuclear structure calculations. 

A crucial role is played by nuclear structure through the one-body density matrix elements (OBDME) $(\Psi_H\,||\,(b_{r}^\dagger\times a_s)^J\,||\;\Phi_A\,)$. This quantity is computed within the framework of the   TD$_{\Lambda}$ approach and   its multiphonon extension  EMPM$_{\Lambda}$. 
To this purpose we use the $NN$ + $NNN$ chiral potential NNLO$_{\rm{sat}}$~\cite{NNLOsat} plus the renomalized G-matrix Nijmegen-F $\Lambda N$ interaction (NF YNG) \cite{Nijmegen} with various Fermi momenta $k_{\rm{F}}$.  
Both approaches are described in the following section.
\section{Nuclear Structure methods}
\label{Structure}
\subsection{Tamm-Dancoff approach for hypernuclei (TD$_\Lambda$)}
\label{TDL}
Let us consider the angular momentum ($J_\lambda$) coupled basis  states 
\begin{equation}
\mid (p \times h^{-1})^\lambda \rangle= (b^\dagger_{p} \times a_h)^\lambda \mid 0 \rangle,
\end{equation}
where $b^\dagger_p$ creates the $\Lambda$ particle $p$  and  $a_h$  a nucleon hole $h^{-1}$  out of the nuclear unperturbed HF ground state $\mid 0 \rangle$.

 Such a basis is adopted to solve the eigenvalue equation
\begin{eqnarray}
&&\langle (p \times h^{-1})^\lambda \mid H \mid \lambda \rangle
= \sum_{p' h'} \Bigl( [ (\epsilon_p - \epsilon_h) - \omega_\lambda] \delta_{pp'} \delta_{h h'}   \nonumber\\
&& +\langle (p \times h^{-1})^\lambda \mid V_{\Lambda N}  \mid (p' \times h'^{-1} )^\lambda \rangle \Bigr) c^\lambda_{p'h'}, 
\end{eqnarray}
where $V_{\Lambda N}$ is the  $\Lambda N$ potential and
\begin{equation}
\omega_{\lambda} = E_\lambda - E_{\sf HF}.
\label{eigTD}
\end{equation}
The eigenstates are
\begin{equation}
\mid \lambda \rangle =  Q^\dagger_\lambda \mid 0 \rangle = \sum_{p h} c^\lambda_{ph} \mid (p \times h^{-1})^\lambda \rangle. 
\end{equation}
The above equation defines $Q^\dagger_\lambda$ as an operator which creates TD$_\Lambda$ hypernuclear phonons. The amplitudes $c^\lambda_{p h}$ yield the TD$_\Lambda$ OBDME
\begin{equation}
c^\lambda_{p h} = \frac{1}{[J_\lambda]} \langle \lambda \parallel  (b^\dagger_{p} \times a_h)^\lambda \parallel 0 \rangle .
\end{equation}

The TD formalism for the nucleonic excitations is exactly the same. We need to replace the $\Lambda N$ potential $V_{\Lambda N}$ with the $NN$ interaction $V_{NN}$ and  the $\Lambda$ with  a nucleon particle. 
We obtain
\begin{equation}
\mid \sigma \rangle =  O^\dagger_\sigma \mid 0 \rangle = \sum_{p h} c^\sigma_{ph} \mid (p \times h^{-1})^\sigma \rangle. 
\end{equation}

\subsection{ Beyond TD$_{\Lambda}$: The EMPM$_{\Lambda}$}
\label{EMPML}
We now construct the basis of orthonormal $n$-phonon states $\mid \beta_n \rangle$ out of the redundant set
 \begin{equation}
 \mid \lambda \alpha_n \rangle =  Q^\dagger_\lambda \mid \alpha_n \rangle,
 \end{equation}
where $\alpha_n$ is an $n$-phonon ($n=1,2,...$) state describing nuclear excitations and is composed of $n$ nuclear TD phonons.
     
 We first extract from the redundant set a basis of linearly independent (but not orthogonal) states  $\mid \lambda \alpha_n \rangle$ through the Cholesky decomposition method. We then start with  the equations of motion
 \begin{equation}
\langle \alpha_n \mid [Q_\lambda,H] \mid \beta _n\rangle=  
(E_{\beta_n} - E_{\alpha_n}) \langle  \lambda \alpha_n \mid \beta_n \rangle,
\label{EoM}
\end{equation}
where $Q_\lambda$ is the adjoint of the TD$_\Lambda$ phonon creation operator.
 
After expanding the commutator and performing additional manipulations, we get
the generalized eigenvalue equations
\begin{equation}
\sum_{jk} \Bigl({\cal H} ^{\beta_n}_{ik}  - E_{\beta_n} \delta_{ik} \Bigr)
{\cal D}^{\beta_n}_{kj}C^{\beta_n}_{j} =  0.
\label{eign}
\end{equation}
Here
\begin{eqnarray}
{\cal H}_{ik}^{\beta_n} =
{\cal H} ^{\beta_n}_{\lambda \alpha_n \lambda" \alpha"_n}  = \nonumber\\
(E_\lambda  + E_{\alpha_n} ) \delta_{\lambda \lambda"} \delta_{\alpha _n\alpha"_n} 
+ {\cal V}^{\beta_n}_{\lambda \alpha_n \lambda" \alpha"_n},
\label{Hn}
 \end{eqnarray}
 where  ${\cal V}^{\beta_n}_{\lambda \alpha_n \lambda" \alpha"_n}$ 
defines the phonon-phonon interaction, and
\begin{equation}
\label{D}
{\cal D}^{\beta_n}_{ij} =  {\cal D}^{\beta_n}_{\lambda \alpha_n \lambda' \alpha'_n}= \langle \lambda ' \alpha'_n \mid \lambda \alpha_n \rangle
\end{equation}
is the overlap or metric matrix which preserves  the Pauli principle. The expressions for  ${\cal D}$ and ${\cal V}$  can be found, for instance, in Ref. \cite{Bianco12}.

The $n$-phonon eigenstates so obtained have the form
\begin{equation}
\mid \beta_n \rangle = \sum_{\lambda \alpha_n}  C_{\lambda \alpha_n}^{\beta_n} \mid \lambda \alpha_n \rangle.
\label{nstate}
\end{equation}
The iteration of the procedure up to an arbitrary $n$ produces a set of states which, added to the 
TD$_{\Lambda}$ states $\{\mid \beta_0\rangle\} = \{\mid \lambda \rangle\}$, form an orthonormal basis $\{\mid \beta_n \rangle\}$ ($n=0,1,2,3,...$).

Such a basis  is used for constructing and solving the eigenvalue problem  in the full space
\begin{equation}
\label{eigfull}
\sum_{ \beta_n \beta_{n'}} \Bigl((E_{\beta_n} - {\cal E}_\nu) \delta_{\beta_n \beta_{n'}} + {\cal V}_{\beta_n \beta_{n'}} \Bigr){\cal C}^{\nu}_{\beta_{n'}} = 0,
\end{equation}
where ${\cal V}_{\beta_n \beta_{n'}}  = 0$ for  $n' = n$. 

The solution yields for the hypernuclei the eigenvectors
\begin{eqnarray}
\label{Psifull}
\mid \Psi_\nu \rangle = \sum_{n,\beta_n}  {\cal C}_{\beta_n}^\nu \mid \beta_n \rangle =\nonumber\\
\sum_{\beta_0}  {\cal C}_{\beta_0}^\nu \mid \beta_0\rangle + \sum_{\beta_1}{\cal C}_{\beta_1}^\nu \mid \beta_1 \rangle + 
\sum_{\beta_2}{\cal C}_{\beta_2}^\nu \mid \beta_2\rangle\,+\,\dots
 \end{eqnarray}
The same method yields for the ground state of the target nuclei
\begin{eqnarray}
\label{Phifull}
\mid \Phi_0 \rangle = \sum_{n,\alpha_n}  {\cal R}_{\alpha_n} ^0 \mid \alpha_n \rangle= \nonumber\\
= {\cal R}_0 ^0 \mid  0 \rangle  +  \sum_{\alpha_2} {\cal R}_{\alpha_2} ^0 \mid \alpha_2 \rangle + \dots.
\end{eqnarray}
The absence of the one-phonon components are to be noticed. It is an effect of the self-consistent HF basis adopted.
The OBDME are given by 
 \begin{eqnarray}
 \langle \Psi_\nu \parallel  (b^\dagger_{r} \times a_s)^\lambda \parallel \Phi_0 \rangle = \nonumber\\ 
 \sum_{n,  \alpha_n,\beta_n} {\cal C}_{\beta_{n}}^\nu {\cal R}_{\alpha_n}^0
 \langle \beta_{n}  \parallel   (b^\dagger_{s} \times a_r)^\lambda \parallel  \alpha_n \rangle.
 \end{eqnarray} 
 
 We consider the restricted space spanned by $\mid \beta_0 \rangle =  \mid \lambda \rangle$  and  $\mid \beta_1 \rangle$. 
Thus the OBDME are simply
 \begin{equation}
 \langle \Psi_\nu \parallel  (b^\dagger_{r} \times a_s)^\lambda \parallel 0 \rangle = \sum_\lambda {\cal C}_{\lambda}^\nu \langle \lambda  \parallel   (b^\dagger_{p} \times a_h)^\lambda \parallel  0 \rangle \delta_{rp} \delta_{sh}.
 \end{equation} 
   It is important to notice that  $\langle \beta_1 \parallel   (b^\dagger_{p} \times a_h)^\lambda \parallel  0 \rangle = 0$.  In order to get an additional contribution and, consequently, additional peaks in the cross section, we should enlarge the space so as to include  the two-phonon basis states $\mid \alpha_2 \rangle$ for the parent nuclei and/or hypernuclei. We would get contributions from
   \begin{equation}
   \label{rho1}
   \langle \beta_1 \parallel  (b^\dagger_{r} \times a_s)^\lambda \parallel  \alpha_2 \rangle \propto \langle  \lambda' \parallel  (b^\dagger_{p} \times a_{p'})^\lambda \parallel  \sigma \rangle \delta_{rp} \delta_{sp'}
         \end{equation}
         and
   \begin{equation}
   \label{rho2}
   \langle \beta_2 \parallel  (b^\dagger_{r} \times a_s)^\lambda \parallel  \alpha_2 \rangle \propto \langle \beta_0= \lambda \parallel  (b^\dagger_{p} \times a_h)^\lambda \parallel  0 \rangle.
         \end{equation}
\section{Calculation details}

The Hamiltonian has the structure
\begin{equation}
 H = T_{\rm{intr}} + V  = T_{N} + T_{\Lambda} - T_{\rm{c.m.}} + V_{\rm{sat}} + V_{\Lambda N} . 
 \label{Hyper-Hamil}
\end{equation}
The intrinsic kinetic term $T_{\rm{intr}}$ is obtained by subtracting the center of mass term $T_{\rm{c.m.}}$ from the kinetic terms of nucleons ($T_{N}$) and $\Lambda$ ($T_{\Lambda}$), $V_{\rm{sat}}$ is the NNLO$_{\rm{sat}}$ potential \cite{NNLOsat} which includes the $NN$ and $NNN$ interactions. The $NNN$ component is fully taken into account in generating the HF basis, while it is truncated at the normal order two-body level in TD (TD$_\Lambda$) and the EMPM (EMPM$_{\Lambda}$). The residual $NNN$ interaction does not enter into the TD (TD$_\Lambda$), which is therefore unaffected by the truncation. 

The  G-matrix derived from the Nijmegen-F interaction  is parametrized as a sum of Gaussian-like terms \cite{Nijmegen} 
\begin{equation}
    V_{\Lambda N} = \sum^{3}_{i=1} (a_i + b_i k_{\rm{F}} + c_i k^2_{\rm{F}} )\,\rm{exp}(-r^2/\beta^2_{i})
\end{equation}
and is used for various values of the Fermi momentum $k_{\rm{F}}$. The $a_i$, $b_i$ and $c_i$ coefficients are given in \cite{Nijmegen}. 

Using the above potentials we have generated a HF basis for $\Lambda$ as well as for the nucleons from a harmonic oscillator (HO) space sufficiently large in order to reach  convergence with respect to the
HO frequency.
It is sufficient to use a HO basis with a number of major shells up to $N_{\rm{max}} = 10$ for $p$-shell hypernuclei and $N_{\rm{max}} = 12$ for $sd$-shell hypernuclei.  The EMPM$_{\Lambda}$ is numerically more costly. Therefore, in $^{12}_{~\Lambda}$B and $^{16}_{~\Lambda}$N we restricted our space to $N_{\rm{max}} = 8$ for the proton and neutron levels, while for $\Lambda$ we took into account levels up to the $sd$-shell. In $^{28}_{~\Lambda}$Al, $^{40}_{~\Lambda}$K, and $^{48}_{~\Lambda}$K, the EMPM$_{\Lambda}$ calculation is unreachable even for $N_{\rm{max}}=8$. Thus,  we have generated the  HF basis for $N_{\rm{max}} = 12$ in order to obtain fully convergence for the nucleon  s.p. states and then used a subset of such a basis corresponding to a HO space up to  $N_{\rm{max}} = 4$. The space for $\Lambda$ is up to the $sd$-shell.  
%
\section{Results}
The cross sections were calculated in various kinematics using the elementary amplitudes BS3 \cite{SB18} and SLA \cite{SLA} in the optimum on-shell approximation \cite{fermi}.  We also considered the frozen-proton approximation in order to relate our findings to the results by Motoba {\it et al.} \cite{Motoba2012}. 
The kaon distortion was treated in the same manner as in our previous calculations~\cite{fermi}. Therefore, the differences with respect to the previous results are to be ascribed to the different methods adopted here to compute the transition densities OBDME and to the different forms of the $YN$ interaction.
The peaks were smoothed through Gaussians with a uniform width deduced from the experimental data or consistent with the width anticipated in planned experiments. The experimental background was not taken into account. 
%
\subsection{The $^{12}_{~\Lambda}$B hypernucleus}

We used the NF YNG  interaction with $k_{\rm{F}}$= 1.1 fm$^{-1}$ to compute
the binding energy of $\Lambda$. This was extracted  through the approximate formula 
\begin{equation}
-B_{\Lambda}  \simeq {\cal E}_{\sf g.s.} +\epsilon^{\rm{p}}_{\rm{F}},
\label{Binding}
\end{equation}
where ${\cal E}_{\sf g.s.} $ is  the  lowest   TD$_{\Lambda}$ (Eq. \ref{eigTD}) or EMPM$_{\Lambda}$ (Eq. \ref{eigfull}) eigenvalue and  $\epsilon ^{\rm{p}}_{\rm{F}}$ is the energy of the last occupied proton orbit. We get $B_{\Lambda}^{\rm{(TD)}} = 10.37$ MeV,  smaller than the empirical binding energy 
$B^{\rm{(exp)}} = 11.37\pm0.06$ MeV \cite{GHM2016,AP2019}. The EMPM${_\Lambda}$  value  is larger instead ($B_{\Lambda}^{\rm{(EM)}} = 12.88$ MeV). 

\begin{figure}[hbt]
\begin{center}
\includegraphics[width=0.359\textwidth,angle=270]{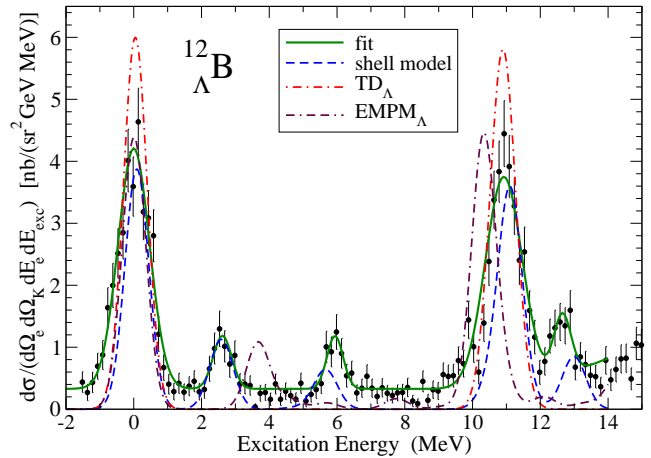}
\end{center}
\caption{ The TD$_{\Lambda}$ (dash-dot line), and EMPM$_\Lambda$ (double-dash-dot line)  spectra of $^{12}_{~\Lambda}$B are compared  with the data \cite{AP2019} and an empirical shell-model calculation \cite{AP2019}. The calculations were performed using the BS3 elementary amplitude. The solid line is the fit to the data. The FWHM used in the calculation is 820 keV.}
\label{spectrum_L12Bx}
\end{figure}

 In Fig. \ref{spectrum_L12Bx}, the TD$_\Lambda$  and EMPM$_\Lambda$ energy distributions of the cross section are compared  with the data produced by the E94-107 experiment~\cite{AP2019} and analyzed within a shell model approach~\cite{AP2019}. Such a calculation used a Woods-Saxon s.p. basis and an effective $YN$ interaction fitted to the $\gamma$-ray spectroscopic data of the $p$-shell hypernuclei~\cite{JohnM}.    In Ref.~\cite{fermi}, we analyzed the same data. To this purpose,  we took the shell model OBDME from Ref.~\cite{AP2019},  but replaced the frozen-proton approximation  adopted there with  the  optimum on-shell approximation. Such  a replacement, complemented with other minor changes, has improved the agreement with the experimental data. In fact, both magnitudes and energy distribution of the transition strengths are well reproduced. 

The  TD$_{\Lambda}$ and EMPM$_\Lambda$ calculations were performed  using a HF basis and they are parameter free except for  the Fermi momentum  $k_{\rm{F}}$. This parameter was fixed   so as to reproduce the empirical energy gap between the   $(0p_{3/2})_\Lambda$ and $(0s_{1/2})_\Lambda$ states and resulted to be $k_{\rm{F}}= 1.1$ fm$^{-1}$.
TD$_{\Lambda}$ reproduces the two main peaks. Though slightly too high, they fall at the correct energies. The low and high energy peaks correspond to the  $(0s_{1/2})_\Lambda- (0p_{3/2})^{-1}_N$ and $(0p_{3/2})_\Lambda-(0p_{3/2})^{-1}_N$ excitations.  
TD$_{\Lambda}$  misses the observed strength in between. The coupling to the nuclear core  TD phonons, within the EMPM$_\Lambda$, induces a damping of the two main peaks consistently with the data,  but also a slight downward energy shift
of the high energy peak. It produces also several transitions comparable  in number with those observed experimentally. However, they are generally too weak and not always at the correct energies. It seems therefore necessary to enlarge the multiphonon
space so as to include the two-phonon (2p-2h) configurations. As illustrated in Eqs. (\ref{rho1}) and (\ref{rho2}) this additional basis is expected to enhance the fragmentation of the cross section as well as  the strengths of the existing intermediate transitions.  
%
%

%
\subsection{The $^{16}_{~\Lambda}$N hypernucleus}

As for  $^{12}_{~\Lambda}$B, we used the NF YNG  interaction with $k_{\rm{F}}$= 1.1 fm$^{-1}$. The TD$_\Lambda$  binding energy is $B_{\Lambda}^{\rm{(TD)}} = 12.93$ MeV, smaller than the empirical value $B^{\rm{(exp)}} = 13.76\pm0.16$ MeV \cite{16Otarget}.
The EMPM$_\Lambda$ yields $B_{\Lambda}^{\rm{(EM)}} = 17.19$ MeV, which is too large. It is  necessary to explore if and how such a large  discrepancy can be reduced. One may try  to exploit the sensitivity to the Fermi momentum of the V$_{\Lambda N}$ potential.
 
\begin{figure}[hbt]
\begin{center}
\includegraphics[width=0.359\textwidth,angle=270]{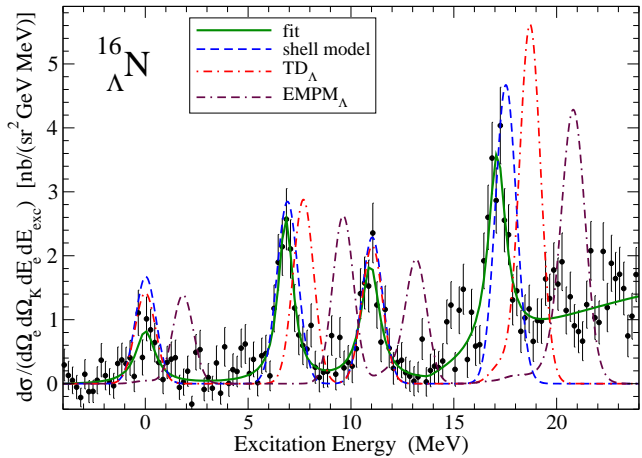}
\end{center}
\caption{The experimental  spectrum of $^{16}_{~\Lambda}$N and its fit, taken from Ref.~\cite{AP2019} are compared with the theoretical cross sections obtained within  shell-model, TD$_\Lambda$, and EMPM$_{\Lambda}$ using the BS3 elementary amplitude. The theoretical lines do not account for the background but the fit includes the background. The FWHM used here is 1177 keV.}
\label{spectrum_L16Nx}
\end{figure}

We computed the cross section by the same shell model approach  adopted for $^{12}_{~\Lambda}$B. As shown in Fig. \ref{spectrum_L16Nx}, the calculation tends to overestimate the magnitude of the peaks, just as in previous investigations \cite{AP2019,fermi}, suggesting the need of enlarging the shell-model space~\cite{AP2019}.  
%
%

Unlike the case of $^{12}_{~\Lambda}$B,  the cross section obtained within TD$_{\Lambda}$,  has the same peak composition of the experimental quantity (Fig. \ref{spectrum_L16Nx}).  The first and third peaks fall at the correct energies. They correspond to the $(0s_{1/2})_\Lambda-(0p_{1/2})^{-1 }_N$  and $(0p_{3/2})_\Lambda-(0p_{1/2})^{-1 }_N$ excitations, respectively. We observe, instead, an upward shift of the second and fourth peaks corresponding to the $(0s_{1/2})_\Lambda-(0p_{3/2})^{-1 }_N$  and $(0p_{3/2})_\Lambda-(0p_{3/2})^{-1 }_N$ configuration, an indication of a possible too large spin-orbit splitting.  
We could bring them to the experimental excitation energies  if we  multiply by a factor $1.2$ the strength of  the $NNN$ component of the  NNLO$_{\rm{sat}}$ potential, while keeping the same  $k_{\rm{F}}$ in the $YN$ potential.

The  EMPM$_\Lambda$  cross section  keeps roughly the peak structure obtained in TD$_\Lambda$. However, it is rigidly shifted upward in energy by 1.83 MeV.    Such a shift finds the following  explanation.    The lowest peak is generated by populating the excited $1^-_1$ state rather than the $0^-_1$ ground state which is shifted down in energy and carries a negligible strength. 
In TD$_\Lambda$, instead, these two states  are almost degenerate, consistently with experimental $\gamma$-ray spectroscopy measurements of the mirror hypernucleus $^{16}_{~\Lambda}$O \cite{Tamura-O16}.   
It would be necessary to enlarge the multiphonon space in order to hopefully clarify the reason of such a deficiency. 
\subsection{The $^{28}_{~\Lambda}$Al hypernucleus}

In the absence of experimental data, we estimated the Fermi momentum $k_{\rm{F}}$  entering the NF YNG potential in the Thomas-Fermi approximation. We have
\begin{equation}
k_{\rm{F}} = \left( \frac{3\pi^2}{2} \langle \rho \rangle \right)^{1/3},
\label{TFA}
\end{equation}
where the average density $\langle \rho \rangle$ is calculated in the average density approximation (ADA)
\begin{equation}
\langle \rho \rangle = \int \rm{d}^3r \rho_{N}(\vec{r}) \rho_{\Lambda}(\vec{r}). 
\end{equation}
Here  $\rho_{N}(\vec{r})$ ($\rho_{\Lambda}(\vec{r})$) is the nucleon ($\Lambda$) density. For $^{28}_{\Lambda}$Al we obtain $k_{\rm{F}} = 1.24$ fm$^{-1}$. 

By using this value we obtain for the binding energy  of $\Lambda$ in its lowest $s$-orbit $B^{\rm{(TD)}}_{\Lambda}= 15.58$ MeV  in TD$_\Lambda$ and %
$B^{\rm{(EM)}}_{\Lambda}= 16.62$ MeV in EMPM$_\Lambda$, both close to the empirical value 
for $^{28}_{~\Lambda}$Si $B^{\rm{(exp)}}_{\Lambda}= 17.2$ MeV~\cite{GHM2016,FG2023}.
%
%
\begin{figure}[htb]
\begin{center}
\includegraphics[width=0.359\textwidth,angle=270]{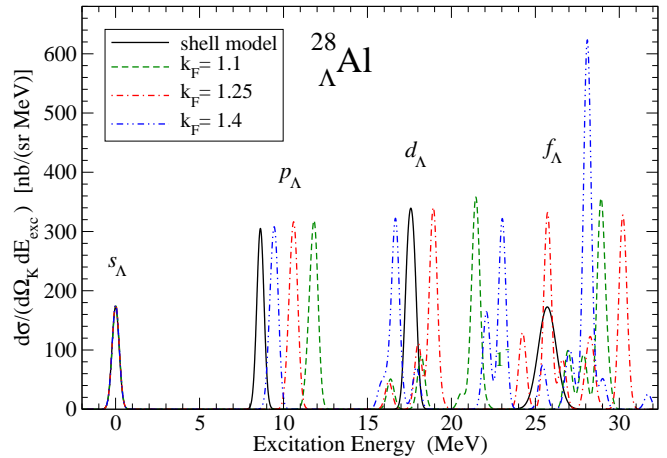}
\end{center}
\caption{The TD$_\Lambda$ electroproduction cross section of $^{28}_{~\Lambda}$Al, computed for three  values of the Fermi momentum $k_{\rm{F}}$ [fm$^{-1}$] entering the NF YNG potential, is compared with the photoproduction cross section obtained within a phenomenological shell model by Motoba {\it et al}~\cite{Motoba2012}.
The peaks are drawn using the FWHM= 500 keV. The labels $s$, $p$, $d$, and $f$  denote the $\Lambda$ orbital quantum numbers  of the   ( $p_\Lambda - h^{-1}$) configurations generating the corresponding peaks.
The differential cross section $d\sigma$ is given in nb/sr (Eq. (\ref{crs})),  while for $^{12}_{~\Lambda}$B and $^{16}_{~\Lambda}$N we plotted the triple-differential cross section in 
nb/(sr$^2$ GeV).}
\label{spectrum_L28Al-1}
\end{figure}

Since the Thomas-Fermi approximation (\ref{TFA}) provides just an estimate of $k_{\sf F}$ we investigate how sensitive are the cross sections to the Fermi momentum.
The  TD$_{\Lambda}$ cross sections computed using  three $k_{\rm{F}}$  values  are plotted in Fig.~\ref{spectrum_L28Al-1} and compared with the photoproduction cross section computed within a phenomenological shell model approach by Motoba {\it et al} \cite{Motoba2012} using the universal  $sd$-shell interaction (USD)~\cite{USD} in the full [$s^4p^{12}(sd)^{12}_{pn}$] model space.
Both calculations made use of the Saclay-Lyon A (SLA) elementary amplitude in the frozen-proton approximation.
Note that the kinematics was $E_\gamma = 1.3$ GeV and $\theta_K = 3^\circ$ in the photoproduction and $E_i = 1.8$ GeV, $E_f = 0.5$ GeV, $\theta_e = 5.4^\circ$, $\theta_{Ke} = 5.1^\circ$, and $\Phi_K = 180^\circ$ in the electroproduction.
Since the photon virtuality given by $Q^2 = -q^2_\gamma = 0.008$~(GeV/c)$^2$ is very small, the  two approaches can be regarded as almost equivalent.
 
The cross section generated by the ground state doublet is practically the same in both approaches. Note the extreme sensitivity of the energy distribution of the peaks to $k_{\rm{F}}$ is to be noticed.
The spectrum tends to be more and more compact as $k_{\rm{F}}$ increases. However, whatever $k_{\rm{F}}$ value is adopted, only a qualitative partial consistency between the shell-model and TD$_\Lambda$ results can be achieved.
  
The very high peak  at 28.1 MeV  which exceeds  300 nb/sr is obtained  for  $k_{\rm{F}}=1.4$ fm$^{-1}$  and originates from  the envelope of many closely packed  transitions. 
The main contribution comes from the states 4$^+$ , 7$^+$, and 5$^+$  at E= 27.979 MeV, E= 28.122 MeV,  and E= 28.142 MeV, respectively.  
These states are mostly populated through the transition from  the  0$d_{5/2}$ proton to the final  1$d_{3/2}$ (4$^+$), 1$d_{5/2}$ (7$^+$), and 0$g_{9/2}$ (5$^+$) $\Lambda$ orbits. The peak at $\approx$ 23 MeV,  whose magnitude is about half,  is produced by the transitions to the 6$^-$  and 5$^-$  states, obtained by creating  a $(0f_{7/2})_\Lambda$ and $(0f_{5/2})_\Lambda$ at the expense of a $0d_{5/2}$ proton. 

%
\subsection{The $^{40}_{~\Lambda}$K hypernucleus}
%
%
%
\begin{figure}[hbt]
\begin{center}
\includegraphics[width=0.359\textwidth,angle=270]{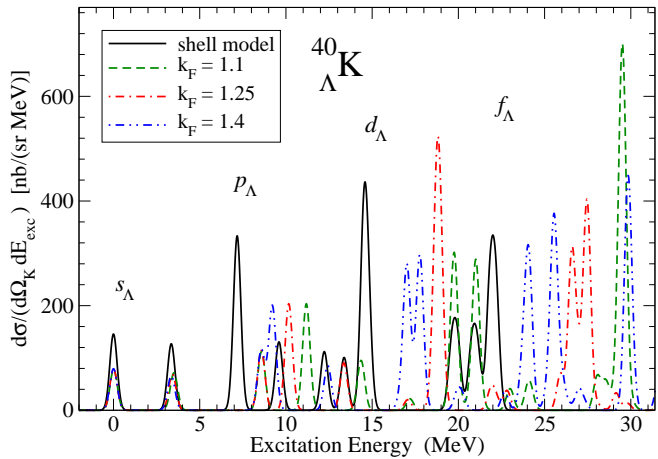}
\end{center}
\caption{The same as in Fig.~\ref{spectrum_L28Al-1} but for $^{40}_{~\Lambda}$K.}
\label{spectrum_L40K-1}
\end{figure}

In $^{40}_{~\Lambda}$K, not only the position, as in  $^{28}_{~\Lambda}$Al, but also the height of the TD$_\Lambda$ peaks changes considerably with $k_{\rm{F}}$ (Fig.~\ref{spectrum_L40K-1}). Also for this system, the agreement with the shell model spectrum is only partial for any $k_{\rm{F}}$.
 
In this stage, it may be worthwhile to make some predictions for the spectrum of $^{40}_{~\Lambda}$K in view of the planned JLab E12-15-008 experiment~\cite{E12-15-008}. For this purpose, we used the following kinematics of the experiment~\cite{Nue}: $E_i= 2.24$ GeV, $E_f= 0.74$ GeV, $\theta_e= 8^\circ$, $\theta_{Ke}= 11^\circ$, and $\Phi_K= 180^\circ$, which generates kinematics of the virtual photon $E_\gamma= 1.5$ GeV,  $\theta_{K\gamma}= 7.1^\circ$, $Q^2= 0.032$ (GeV/c)$^2$, and $\epsilon= 0.591$. We used the FWHM = 800 keV which we deem to be suitable for  the planned experiment. 

\begin{figure}[htb]
\begin{center}
\includegraphics[width=0.359\textwidth,angle=270]{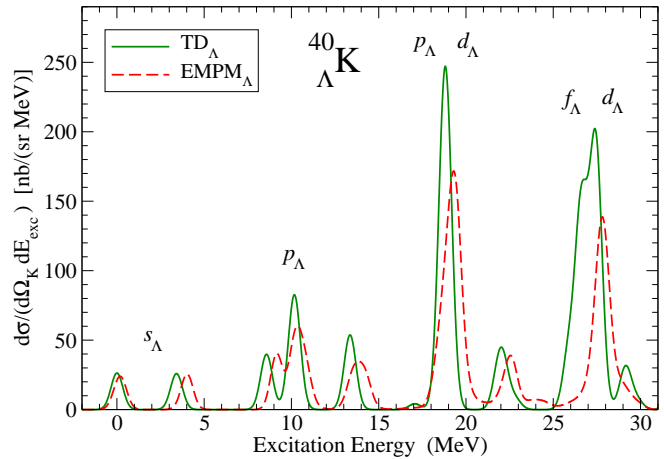}
\end{center}
\caption{The spectrum of $^{40}_{~\Lambda}$K calculated within the TD$_\Lambda$ and EMPM$_{\Lambda}$ approaches with $k_{\sf F}=1.25$ fm$^{-1}$ in kinematics of the E12-15-008 experiment.}
\label{spectrum_L40K-3xx}
\end{figure}

The TD$_\Lambda$ and EMPM$_{\Lambda}$   strength distributions were computed using $k_{\rm{F}}= 1.25$ fm$^{-1}$  and the BS3 amplitude in the optimum on-shell approximation. The two spectra, calculated in kinematics of the E12-15-008 experiment,  are very close (Fig.~\ref{spectrum_L40K-3xx}). 
The number of peaks is comparable in both cases. One may only notice the damping effect  induced by the coupling  of TD$_\Lambda$  to the  nuclear core excitations, described by TD phonons, which reduces systematically the heights of EMPM$_{\Lambda}$ peaks.
Such a coupling induces also a slight upward displacement of the peaks which increases with energy. 
For instance, the first excited state 2$^+$, nearly degenerate with the ground state 1$^+$ in TD$_\Lambda$, is shifted up to 0.269 MeV in EMPM$_\Lambda$. 
The strongly populated EMPM$_\Lambda$ states 1$^+$ and 3$^+$ are $ \approx$ 0.6 MeV above the corresponding TD$_\Lambda$ states at 3.42 and 8.57 MeV, respectively. 

%
%

The coupling to the nuclear core phonons induces an increment of 1.35 MeV to  the binding energy  of $\Lambda$. Using Eq. (\ref{Binding}), we get  for $k_{\rm F}=$ 1.25 fm$^{-1}$  $B^{\rm{(TD)}}_{\Lambda}$ = 18.62 MeV in TD$_\Lambda$ and $B^{\rm{(EM)}}_{\Lambda}$ = 19.97 MeV in the EMPM$_\Lambda$ 
where the former agrees very well with the value $B_{\Lambda}$ = 18.70 MeV calculated from the 
$\Lambda$-nucleus optical potential~\cite{FG2023}. 
The enhanced value of $B^{\rm{(EM)}}_{\Lambda}$ may indicate a need for elaborating the EMPM$_\Lambda$  method as we have mentioned it in the case of the $^{12}$C and $^{16}$O targets but this value is not so bad if we compare it with another empirical value $\approx$ 19.7 MeV obtained from a Woods-Saxon potential in Fig.~11 of Ref.~\cite{GHM2016}.  
%
%
\begin{figure}[htb]
\begin{center}
\includegraphics[width=0.359\textwidth,angle=270]{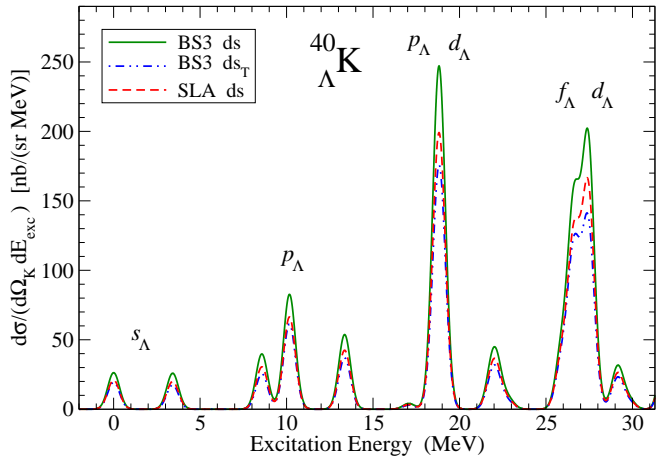}
\end{center}
\caption{Electroproduction (ds) cross sections of $^{40}_{~\Lambda}$K  computed within TD$_\Lambda$  using the BS3 and SLA elementary amplitudes in the optimum on-shell approximation.  The photo-production (ds$_T$), computed with the BS3 amplitude, is also shown. The used FWHM is 800 keV.} 
\label{spectrum_L40K-4}
\end{figure}

Since the target $^{40}$Ca is a closed-shell nucleus, we expect reliable predictions from the TD$_\Lambda$ (particle-hole) approach as was the case
of the reactions using $^{16}$O  as target (Fig.~\ref{spectrum_L16Nx}).  
As shown in  Fig.~\ref{spectrum_L40K-4}, the TD$_\Lambda$ spectrum is affected by the elementary amplitude adopted. 
The BS3 amplitude yields significantly larger cross sections with respect to SLA (see also Table~\ref{table_40Ca}).  
However, the above differences  are smaller than the differences between the photoproduction, given roughly by  ds$_{\rm T}$, and the electroproduction cross sections obtained by using the BS3 amplitude.
Although the photon virtuality $Q^2$ is quite small here, contributions from the longitudinal response functions to the full cross section are important.

The hypernucleus states populated in kinematics of E12-15-008 with cross sections larger than 5 nb/sr are displayed in Table~\ref{table_40Ca}. 
The proton and $\Lambda$ s.p. states with the dominant transition given by OBDME in Eq.~(\ref{amplitude-3}) are also shown. 
The first five states are deeply bound with the $\Lambda$ predominantly in the $0s_{1/2}$ orbit. Consistently with the results, in Fig.~4 of Ref.~\cite{Cohen} one may notice  the deeply-bound states $3^+$ and $2^+$ with the $\Lambda$ in the $0s_{1/2}$ orbit and the strongly populated substitutional states 3$^+$ at $E_x = 18.917$ MeV and  the 1$^+$ at  $E_x = 21.862$ MeV, produced upon replacing a proton with   the $\Lambda$ in the same orbit. Similar results were obtained within the shell model \cite{Motoba2012}.  
%
%
\begin{table}[hbt]
\begin{tabular}{ccccrr}
\hline 
$E_x$\ \ &\ J$^P$ & \multicolumn{2}{c}{s.p. states} &  
                       \multicolumn{2}{c}{cross sections} \\
(MeV)    &        &      p      &  $\Lambda$  &   SLA  & BS3  \\ 
 \hline
  0.000 & $1^+$ &\  $0d_{3/2}$ &\  $0s_{1/2}$ &\    4.47 &   7.55\\
  0.004 & $2^+$ &\  $0d_{3/2}$ &\  $0s_{1/2}$ &\   13.43 &  14.84\\
  3.417 & $1^+$ &\  $1s_{1/2}$ &\  $0s_{1/2}$ &\   17.02 &  21.77\\
  8.570 & $3^+$ &\  $0d_{5/2}$ &\  $0s_{1/2}$ &\   17.34 &  24.53\\
  8.578 & $2^+$ &\  $0d_{5/2}$ &\  $0s_{1/2}$ &\    8.63 &   9.36\\
 10.161 & $2^-$ &\  $0d_{3/2}$ &\  $0p_{1/2}$, $0p_{3/2}$ &\ 18.54 & 28.12\\
 10.164 & $3^-$ &\  $0d_{3/2}$ &\  $0p_{3/2}$ &\   36.49 &  40.42\\
 13.351 & $1^-$ &\  $1s_{1/2}$ &\  $0p_{1/2}$, $0p_{3/2}$ &\ 17.56 & 17.82\\
 13.361 & $2^-$ &\  $1s_{1/2}$ &\  $0p_{3/2}$ &\   17.56 &  23.69\\
 18.511 & $2^+$ &\  $0d_{3/2}$ &\  $1s_{1/2}$ &\    6.07 &   6.41\\
 18.712 & $3^-$ &\  $0d_{5/2}$ &\  $0p_{1/2}$, $0p_{3/2}$ &\ 29.32 & 30.92\\
 18.740 & $4^-$ &\  $0d_{5/2}$ &\  $0p_{3/2}$ &\   50.10 &  69.25\\
 18.917 & $3^+$ &\  $0d_{3/2}$ &\  $0d_{3/2}$ &\   30.20 &  43.14\\
 18.924 & $4^+$ &\  $0d_{3/2}$ &\  $0d_{5/2}$ &\   49.51 &  54.19\\
 18.970 & $2^-$ &\  $0d_{5/2}$ &\  $0p_{1/2}$, $0p_{3/2}$ &\  4.78 &  6.13\\
 21.862 & $1^+$ &\  $1s_{1/2}$ &\  $1s_{1/2}$ &\   14.87 &  18.07\\
 22.128 & $2^+$ &\  $1s_{1/2}$ &\  $0d_{3/2}$ &\    6.64 &   6.54\\
 22.138 & $3^+$ &\  $1s_{1/2}$ &\  $0d_{5/2}$ &\    6.95 &   9.36\\
 25.887 & $2^-$ &\  $0d_{3/2}$ &\  $1p_{1/2}$ &\   10.12 &  14.67\\
 25.905 & $3^-$ &\  $0d_{3/2}$ &\  $1p_{3/2}$ &\   18.31 &  19.35\\
 26.577 & $4^-$ &\  $0d_{3/2}$ &\  $0f_{5/2}$ &\   34.78 &  48.07\\
 26.586 & $5^-$ &\  $0d_{3/2}$ &\  $0f_{7/2}$ &\   52.12 &  56.21\\
 26.947 & $2^+$ &\  $0p_{1/2}$ &\  $0p_{3/2}$ &\    9.06 &   9.78\\
 27.031 & $3^+$ &\  $0d_{5/2}$ &\  $1s_{1/2}$ &\    9.05 &  11.82\\
 27.345 & $4^+$ &\  $0d_{5/2}$ &\  $0d_{5/2}$ &\    6.93 &   7.39\\
 27.463 & $4^+$ &\  $0d_{5/2}$ &\  $0d_{3/2}$ &\   43.34 &  44.47\\
 27.475 & $5^+$ &\  $0d_{5/2}$ &\  $0d_{5/2}$ &\   67.84 &  91.17\\
 29.139 & $1^-$ &\  $1s_{1/2}$ &\  $1p_{1/2}$ &\   10.82 &  10.60\\
 29.148 & $2^-$ &\  $1s_{1/2}$ &\  $1p_{3/2}$ &\   10.65 &  13.44\\
\hline
\end{tabular}
\caption{The cross sections predicted in electroproduction of $^{40}_{~\Lambda}$K for selected hypernucleus states given by $E_x$ and J$^P$. The calculations were done in the TD$_\Lambda$ formalism with the NF YNG interaction and $k_{\rm{F}}= 1.25$ fm$^{-1}$ for two elementary amplitudes SLA and BS3 in the optimum on-shell approximation. The s.p. states of the proton (p) and $\Lambda$, denoted by $r$ and $s$ in Eq.~(\ref{amplitude-3}), for the dominant OBDME are shown. The differential cross section is in nb/sr.}
\label{table_40Ca}
\end{table}  

%
%
\subsection{The $^{48}_{~\Lambda}$K hypernucleus}
In $^{48}_{~\Lambda}$K, the  coupling of TD$_\Lambda$ to the nuclear core excitations within the  EMPM$_\Lambda$ has the effect of incrementing the binding of $\Lambda$ by 1.51~MeV, to be added to the binding obtained with TD$_\Lambda$.
For $k_{\rm{F}} = 1.25$ fm$^{-1}$, $B^{\rm{(TD)}}_{\Lambda} = 20.01$~MeV, which is fairly close to 
the value $B_{\Lambda} = 19.78$ MeV obtained in the calculations with the $\Lambda$-nucleus optical potential~\cite{FG2023}. 

\begin{figure}[htb]
\begin{center}
\includegraphics[width=0.359\textwidth,angle=270]{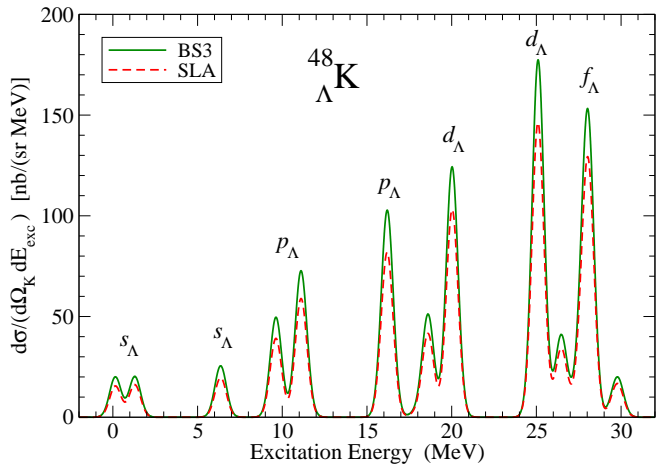}
\end{center}
\caption{ TD$_\Lambda$ $^{48}_{~\Lambda}$K spectrum computed with the BS3 and SLA elementary amplitudes in the optimum on-shell approximation and kinematics of the E12-15-008 experiment.}
\label{spectrum_L48K-4new}
\end{figure}

This coupling induces also a damping and an additional fragmentation of the cross section which modifies the shapes of the peaks in the spectrum similarly as for the $^{40}_{~\Lambda}$K in  Fig.~\ref{spectrum_L40K-3xx}. 
Finally, it should be  pointed out  that the magnitude of the peaks  is also affected  by the elementary amplitude~\cite{AP2019,fermi} (Fig. \ref{spectrum_L48K-4new}) as well as by kinematical and other effects~\cite{fermi,HYP2018}. 
The shape of the spectrum differs from the one predicted for $^{40}_{~\Lambda}$K  (Fig.~\ref{spectrum_L40K-4}) and the cross sections are in general smaller.

%
%

%
%
%
\begin{table}[hbt]
\begin{tabular}{ccccrr}
\hline 
$E_x$\ \ &\ J$^P$ & \multicolumn{2}{c}{s.p. states} &  
                       \multicolumn{2}{c}{cross sections} \\
(MeV)    &        &      p      &  $\Lambda$  &   SLA  & BS3  \\ 
 \hline
  0.148 & $1^+$ &\  $1s_{1/2}$ &\  $0s_{1/2}$ &\   13.14 &  16.80\\
  1.294 & $1^+$ &\  $0d_{3/2}$ &\  $0s_{1/2}$ &\    3.41 &   5.67\\
  1.297 & $2^+$ &\  $0d_{3/2}$ &\  $0s_{1/2}$ &\   10.30 &  11.52\\
  6.359 & $3^+$ &\  $0d_{5/2}$ &\  $0s_{1/2}$ &\   11.14 &  15.61\\
  6.364 & $2^+$ &\  $0d_{5/2}$ &\  $0s_{1/2}$ &\    5.56 &   6.10\\
  9.578 & $1^-$ &\  $1s_{1/2}$ &\  $0p_{1/2}$ &\   14.43 &  14.69\\
  9.657 & $2^-$ &\  $1s_{1/2}$ &\  $0p_{3/2}$ &\   17.05 &  23.06\\
 11.068 & $2^-$ &\  $0d_{3/2}$ &\  $0p_{1/2}$, $0p_{3/2}$ &\ 13.19 & 20.63\\
 11.117 & $3^-$ &\  $0d_{3/2}$ &\  $0p_{3/2}$ &\   31.59 &  35.02\\
 15.975 & $3^-$ &\  $0d_{5/2}$ &\  $0p_{1/2}$, $0p_{3/2}$ &\  6.38 &  6.71\\
 16.140 & $3^-$ &\  $0d_{5/2}$ &\  $0p_{1/2}$, $0p_{3/2}$ &\ 19.37 & 20.67\\
 16.217 & $4^-$ &\  $0d_{5/2}$ &\  $0p_{3/2}$ &\   39.09 &  53.65\\
 16.417 & $2^-$ &\  $0d_{5/2}$ &\  $0p_{1/2}$, $0p_{3/2}$ &\  5.32 &  6.65\\
 18.485 & $2^+$ &\  $1s_{1/2}$ &\  $0d_{3/2}$ &\    7.01 &   6.95\\
 18.603 & $3^+$ &\  $1s_{1/2}$ &\  $0d_{5/2}$ &\    8.45 &  11.43\\
 18.671 & $1^+$ &\  $1s_{1/2}$ &\  $1s_{1/2}$ &\   15.58 &  18.92\\
 19.941 & $3^+$ &\  $0d_{3/2}$ &\  $0d_{3/2}$ &\   22.13 &  30.54\\
 20.006 & $3^+$ &\  $0d_{3/2}$ &\  $0d_{5/2}$ &\    7.42 &  10.49\\
 20.067 & $4^+$ &\  $0d_{3/2}$ &\  $0d_{5/2}$ &\   46.85 &  51.12\\
 24.911 & $3^+$ &\  $0d_{5/2}$ &\  $0d_{3/2}$ &\    6.64 &   6.93\\
 24.980 & $4^+$ &\  $0d_{5/2}$ &\  $0d_{3/2}$, $0d_{5/2}$ &\ 28.51 & 28.96\\
 25.065 & $4^+$ &\  $0d_{5/2}$ &\  $0d_{3/2}$, $0d_{5/2}$ &\ 17.88 & 18.90\\
 25.155 & $5^+$ &\  $0d_{5/2}$ &\  $0d_{5/2}$ &\   62.02 &  82.93\\
 25.248 & $3^+$ &\  $0d_{5/2}$ &\  $0d_{5/2}$, $1s_{1/2}$ &\  6.54 &  7.85\\
 26.430 & $1^-$ &\  $1s_{1/2}$ &\  $1p_{1/2}$ &\   10.95 &  10.73\\
 26.441 & $2^-$ &\  $1s_{1/2}$ &\  $1p_{3/2}$ &\   12.14 &  15.29\\
 27.434 & $2^+$ &\  $0p_{1/2}$ &\  $0p_{3/2}$ &\    5.09 &   5.60\\
 27.813 & $2^-$ &\  $0d_{3/2}$ &\  $1p_{1/2}$ &\    4.20 &   5.68\\
 27.820 & $2^-$ &\  $0d_{3/2}$ &\  $1p_{3/2}$ &\    5.06 &   7.27\\
 27.847 & $3^-$ &\  $0d_{3/2}$ &\  $1p_{3/2}$ &\   15.71 &  16.62\\
 28.013 & $4^-$ &\  $0d_{3/2}$ &\  $0f_{5/2}$ &\   28.78 &  40.07\\
 28.077 & $4^-$ &\  $0d_{3/2}$ &\  $0f_{7/2}$ &\    6.14 &   6.43\\
 28.129 & $5^-$ &\  $0d_{3/2}$ &\  $0f_{7/2}$ &\   49.17 &  52.93\\
 29.855 & $3^+$ &\  $0p_{3/2}$ &\  $0p_{3/2}$ &\    8.65 &  10.75\\
\hline
\end{tabular}
\caption{The same as in Table~\ref{table_40Ca} but for the $^{48}_{~\Lambda}$K hypernucleus.}
\label{table_48Ca}
\end{table}

In Table~\ref{table_48Ca}, we show the electroproduction cross sections of $^{48}_{~\Lambda}$K.  The  ground state 0$^+$ cross section is negligible whereas the first excited state 1$^+$ is quite strongly populated. 
This can be easily explained. Since the cross section is determined mainly by the s.p. transition $1s_{1/2} \to 0s_{1/2}^\Lambda$, only the $L=0$ in Eq. (\ref{amplitude-3}) contributes so that the production amplitude is determined by the spin $J = S$.  
In the considered kinematics the $S=1$ elementary amplitude (Eq.~(\ref{amplitude-3a}) ) is more than one order of magnitude larger than the $S=0$ one and, consequently, the cross sections are larger by  two orders of magnitude. 

The order of the deepest states 1$^+$ and 2$^+$ in Table~\ref{table_48Ca}, dominated by the (0s$_{1/2}$)$_\Lambda$ -- (1s$_{1/2}$)$_p$ and (0s$_{1/2}$)$_\Lambda$ -- (0d$_{3/2}$)$_p$   configurations, respectively, is reversed with respect to Table I. This effect is caused by different relative energy gap of the proton 0d$_{3/2}$ and 1s$_{1/2}$ orbitals. In $^{40}$Ca ($\epsilon_p$(0d$_{3/2}$) - $\epsilon_p$(1s$_{1/2}$)) = 3.22 MeV while in $^{48}$Ca ($\epsilon_p$(0d$_{3/2}$) - $\epsilon_p$(1s$_{1/2}$)) = $-1.44$ MeV. 
The orbits 0d$_{3/2}$ and 1s$_{1/2}$ switch their positions as a result of the proton-
neutron interaction in $^{48}$Ca with the neutron excess.

As for $^{40}_{~\Lambda}$K, appreciable differences between the SLA and BS3 cross sections can be observed for some peaks (Fig. \ref{spectrum_L48K-4new}). However, these  differences are smaller than the uncertainties caused by neglecting the contributions from the longitudinal mode of the virtual photon. Also in $^{48}_{~\Lambda}$K,  the high energy substitutional states $1^+$, $3^+$, and $5^+$ are strongly populated and the $1^+$, $2^+$, and $3^+$ states with $\Lambda$ in the $0s_{1/2}$ orbit are deeply bound (Table~\ref{table_48Ca}).
%
\section{Conclusions}
We expanded our previous calculations for the electroproduction of $p$-shell hypernuclei to the production of medium-mass ($sd$-shell) hypernuclei and performed a preliminary  analysis of the results. To this end we adopted new formalisms, TD$_\Lambda$ and its extension EMPM$_{\Lambda}$, in order to study the effect of the hypernuclear structure, determined by the OBDME, on the cross sections computed in DWIA. 

Using the effective Nijmegen $YN$ interaction and two amplitudes for the elementary production in the optimum on-shell and frozen-proton approximations, we analyzed the dependence of the energy spectra on the interaction for different values of the Fermi momentum $k_{\rm{F}}$ and estimated the uncertainties from various inputs  (elementary amplitudes, Fermi motion).

The results for the light hypernuclei $^{12}_{~\Lambda}$B and $^{16}_{~\Lambda}$N were compared with the experimental data as well as with our previous calculations using a phenomenological shell-model.
A reasonable agreement was obtained for $^{16}_{~\Lambda}$N whereas our new results for $^{12}_{~\Lambda}$B agree only for the two main peaks of the spectrum. 
In the medium-mass hypernuclei $^{28}_{~\Lambda}$Al and $^{40}_{~\Lambda}$K, the cross sections were compared with older shell-model results. We give also predictions of spectra and electroproduction cross sections of $^{40}_{~\Lambda}$K and $^{48}_{~\Lambda}$K 
in view of the planned E12-15-008 experiment at JLab. 

Our theoretical spectra and cross sections are quite sensitive to the $k_{\rm{F}}$  values used for the $YN$ potential, and, therefore, may provide a useful tool for a more detailed study of the properties of both effective $YN$ and modern realistic nuclear potentials.

According to the theoretical analysis of  the electroproduction of $^{40}_{~\Lambda}$K and $^{48}_{~\Lambda}$K,  important  contributions  to the cross section should come from the longitudinal mode of the virtual photon suggesting that reliable results can be provided only by the electroproduction calculation. 

In order to provide a more complete information in view of the planned  E12-15-008 experiment, we intend to include other effective $YN$ and the three-body $YNN$   interactions and to improve the description of the structure by performing our EMPM$_\Lambda$ calculation within a larger multiphonon space.    

\section*{ACKNOWLEDGEMENT}
The authors thank Ji\v{r}\'{i} Mare\v{s}, John Millener, and Petr Navr\'{a}til for useful discussions. 
We thank Petr Navr\'{a}til also for providing us with the matrix
elements of the NNLO$_{\rm{sat}}$ potential. 
The work was partly supported by the Czech Science Foundation GACR, Grant No. P203-23-06439S. 
This work is co-funded by EU-FESR, PON Ricerca e Innovazione 2014-2020 - DM 1062/2021. 
P.V. thanks the INFN for financial support. Computational resources were provided by
the CESNET LM2015042 and the CERIT Scientific Cloud
LM2015085, under the program “Projects of Large Research,
Development, and Innovations Infrastructures”. 

%
%
%

%

\begin{thebibliography}{99}
\bibitem{PB} We regret having overlooked these two works in Ref.~\cite{fermi}. 
P. Byd\v{z}ovsk\'y fully assumes responsibility for this omission and apologizes for it.
\end{thebibliography}
\end{document}